\documentclass[a0paper,12pt]{article}
\usepackage[english]{babel}
\usepackage[utf8]{inputenc}
\usepackage[T1]{fontenc}
\usepackage{lmodern}
\usepackage[export]{adjustbox}

\usepackage{url}
\usepackage{hyperref}
\usepackage{graphicx}
\usepackage{indentfirst}
\def\institute{Regional Center of Advanced Technologies and Materials\\
Joint Laboratory of Optics of Palacký University and Institute of Physics AS CR, Faculty of Science, Palacký University, 17. listopadu 12, 771 46 Olomouc, Czech Republic}

\def\Title#1{\begin{center} {\Large #1 } \end{center}}
\def\Author#1{\begin{center}{ \sc #1} \end{center}}
\def\Address#1{\begin{center}{ \it #1} \end{center}}

\newenvironment{Presented}{\begin{quotation} \begin{center} 
             PRESENTED AT\end{center}\bigskip 
      \begin{center}\begin{large}}{\end{large}\end{center} \end{quotation}}

\begin{document}
\begin{titlepage}

\vfill
\Title{Study of the semi-boosted topology in top anti-top quark pair spectra} 
\vfill
\Author{\underline{J.~Pácalt}}
\Address{\institute}
\vfill

\begin{abstract}
The aim of this contribution is to present results of adding the semi-boosted transition region between the resolved and boosted top quark pair reconstruction in the $\ell$ + jets channel. This region is often omitted due to more complex reconstruction, which usually leads to a smaller efficiency. The reconstruction is performed for all three topologies for comparison and combination purposes. Hypothetical particle $Z'$ was used as a probe to study the increase in the number of events in the studied energy region.
\end{abstract}

\vfill
\begin{Presented}
$12^\mathrm{th}$ International Workshop on Top Quark Physics\\
Beijing, People's Republic of China, September 22--27, 2019
\end{Presented}
\vfill
\end{titlepage}
\section{Introduction}
This contribution aims to explore possibilities of adding an almost non-used semi-boosted topology into reconstruction procedures for studies of top anti-top quark pair spectra in the semi-leptonic decay channel ($\ell$+jets). The top anti-top quark pair decay channels are named according to the way the $W$ boson from the decay of top quark decays either to a lepton and a neutrino (leptonic decay), or to a pair of a quark and an anti-quark (hadronic decay). In the semi-leptonic top anti-top decay channel, one $W$ boson decays leptonically and the other one hadronically. Two topologies are commonly used at present, resolved and boosted, but in between them lies a region, where the resolved topology transits to the boosted one and neither of those has a full reconstruction efficiency. This semi-boosted region of energies is often omitted from analyses.\par

\section{Reconstruction of events and topologies}

\begin{figure}
\begin{center}
\begin{tabular}{c}
\includegraphics [valign=c,width=0.95\textwidth] {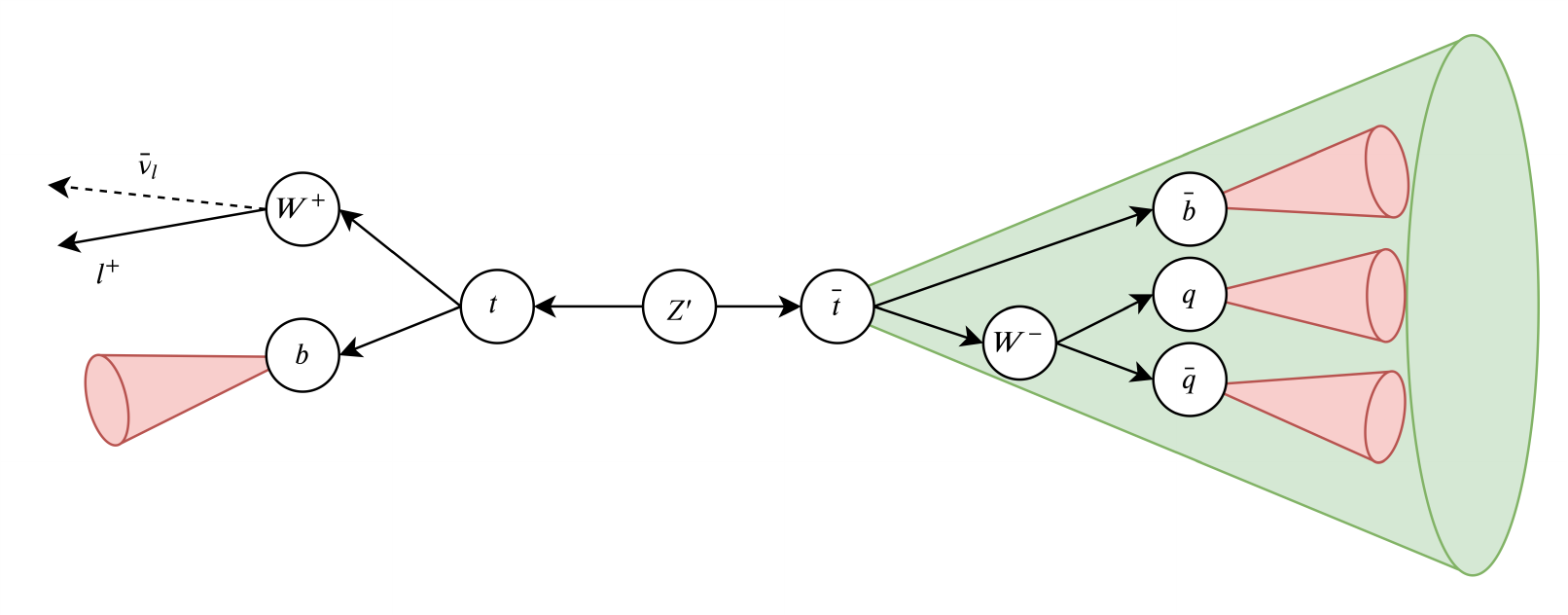} \\
a)\\
\includegraphics [valign=c,width=0.95\textwidth] {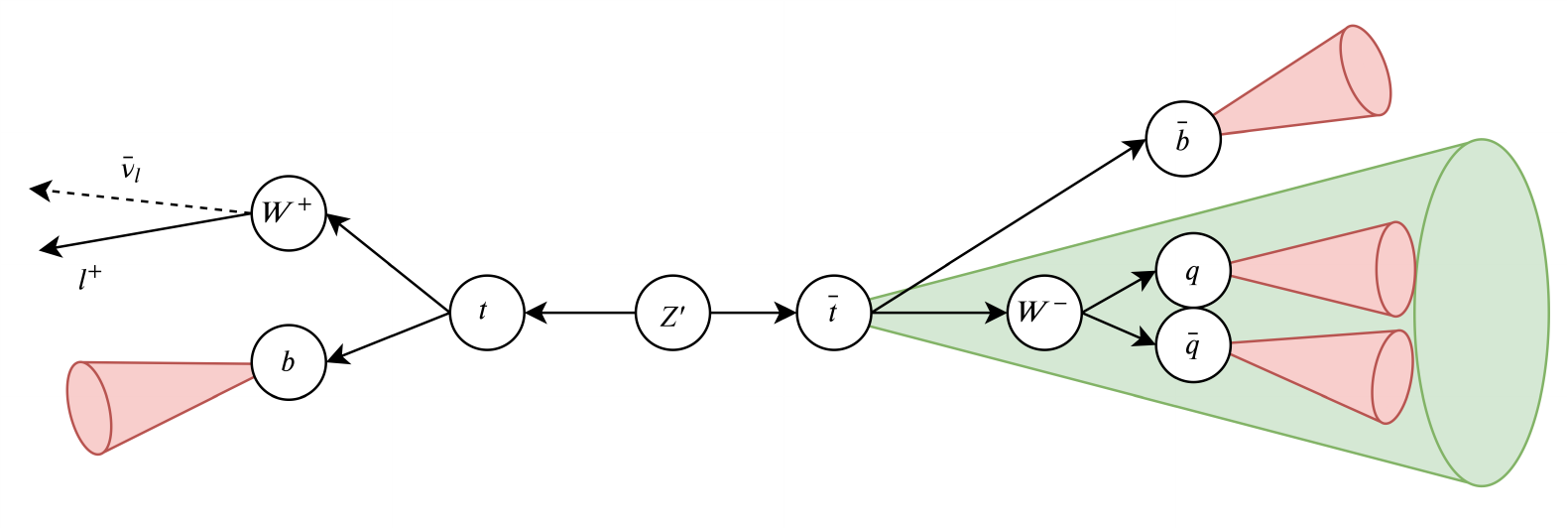}\\
b)\\
\end{tabular}
\begin{tabular}{cc}
\includegraphics [valign=c,width=0.45\textwidth] {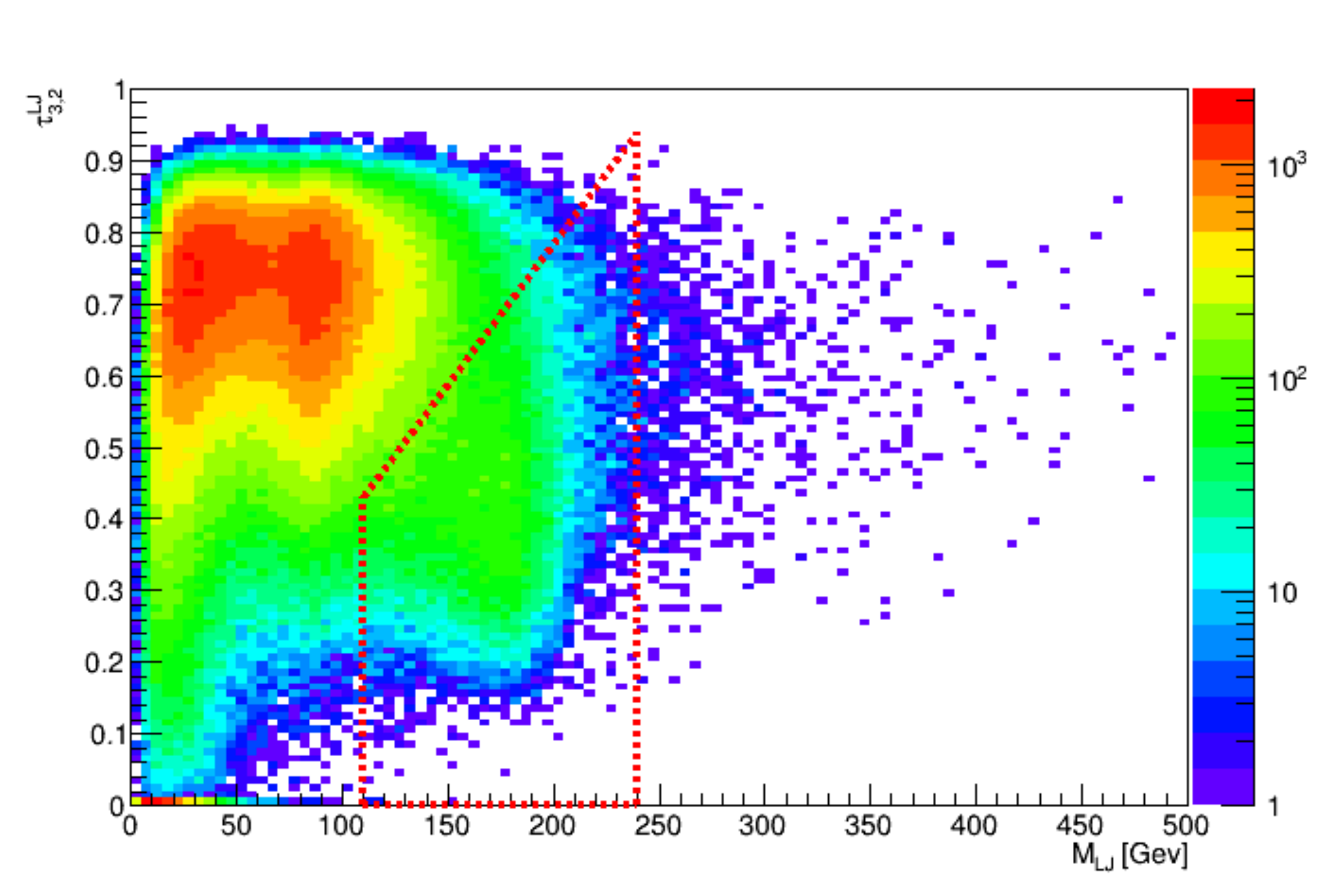} &
\includegraphics [valign=c,width=0.45\textwidth] {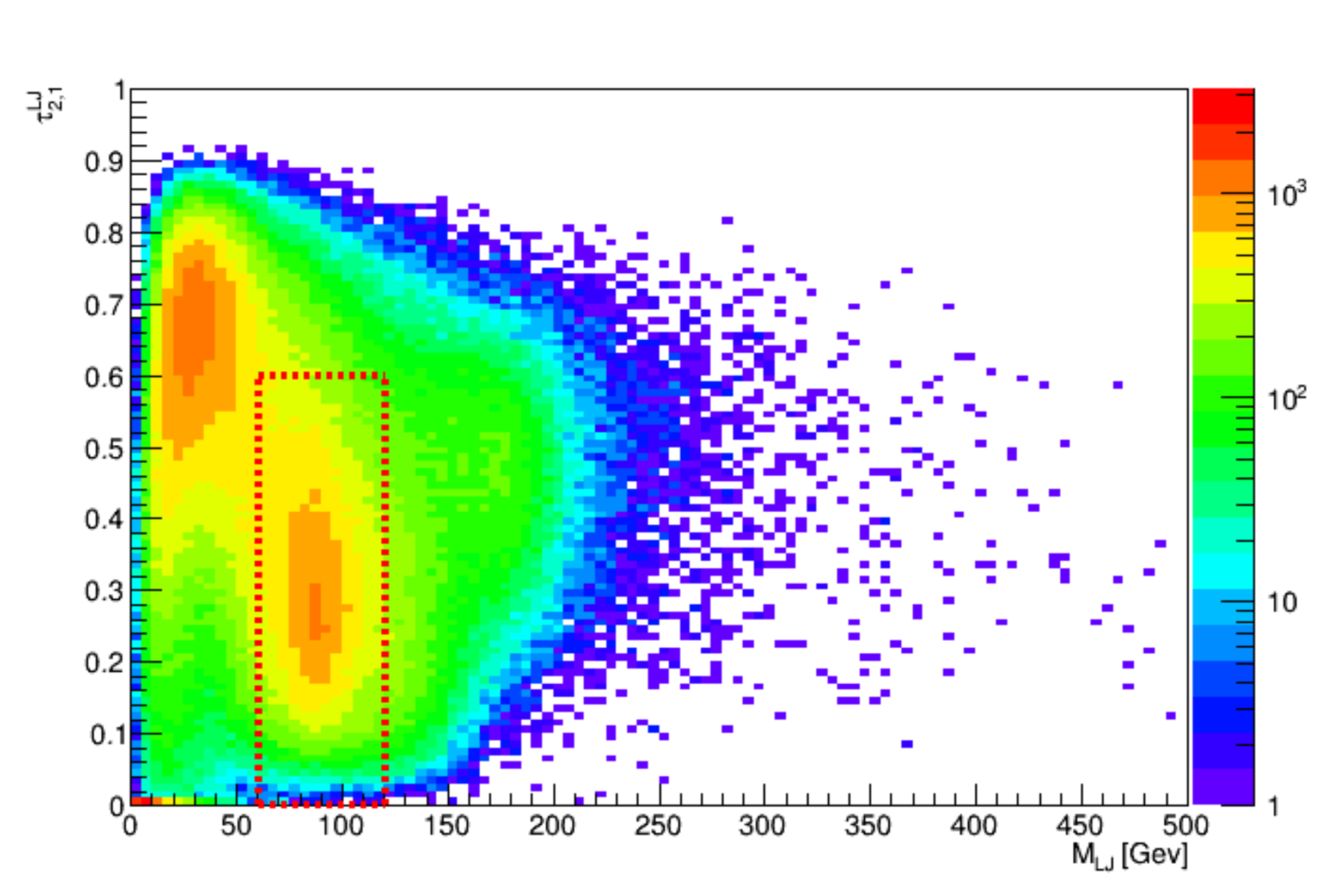}\\
c)&d)\\
\end{tabular}
\caption{Schematic of the studied boosted~(a) and semi-boosted~(b) topologies. Tagging regions of large jets as top quark candidates in the $\tau_{32}$ and large jet mass plane~(c); and \textit{W} boson candidates in the $\tau_{21}$ and large jet mass plane~(d).}
\label{topologies}
\end{center}
\end{figure}

The resolved topology considers all reconstructed elements (jets and leptons) entering the to quark reconstruction angularly separated from each other, in contrary to the boosted topology, which uses collimated products (jets) coming from the hadronically decaying top quark, reconstructed as one object (large jet). The semi-boosted topology is a transition between those two topologies, where the \textit{b}-tagged \footnote{At Large Hadron Collider is the \textit{b}-tagged jet defined as a jet with a secondary vertex, which was identified by the algorithm as a decay of the $B$-hadron inside the jet. The Delphes package uses an efficiency formula from the real detector for the determination of the flavor of quark, this also simulates the misidentification of the tagging algorithm.} jet coming from the hadronically decaying top quark is not merged together with the two jets from the \textit{W} boson decay, which are collimated, or one jet coming from \textit{W} boson decay is separated while the other is collimated with the \textit{b}-tagged jet (semi-boosted mixed topology). Reconstruction of the leptonically decaying top quark is the same for all topologies by using the lepton with high a transverse momentum, $E_{\mathrm{T,miss}}$ (missing energy in transverse plane, which is usually carried away by neutrinos) and the \textit{b}-tagged jet closest to the lepton by imposing the condition $m_{\ell\nu} = m_{\mathrm{W}}$. The situation for the boosted and semi-boosted topologies is shown in Figure \ref{topologies} a) and b), respectively. Tagging of large jets as coming either from the \textit{W} boson or the top quark decay is done by using a substructure variable $\tau_{32}$ \cite{Thaler_2011} for the top quark tagging and $\tau_{21}$ for the \textit{W} boson tagging, and a mass window for the large jet, the chosen regions are shown in Figure \ref{topologies} c) and d), respectively.

\begin{figure}
\begin{center}
\begin{tabular}{cc}
\includegraphics [valign=c,width=0.45\textwidth] {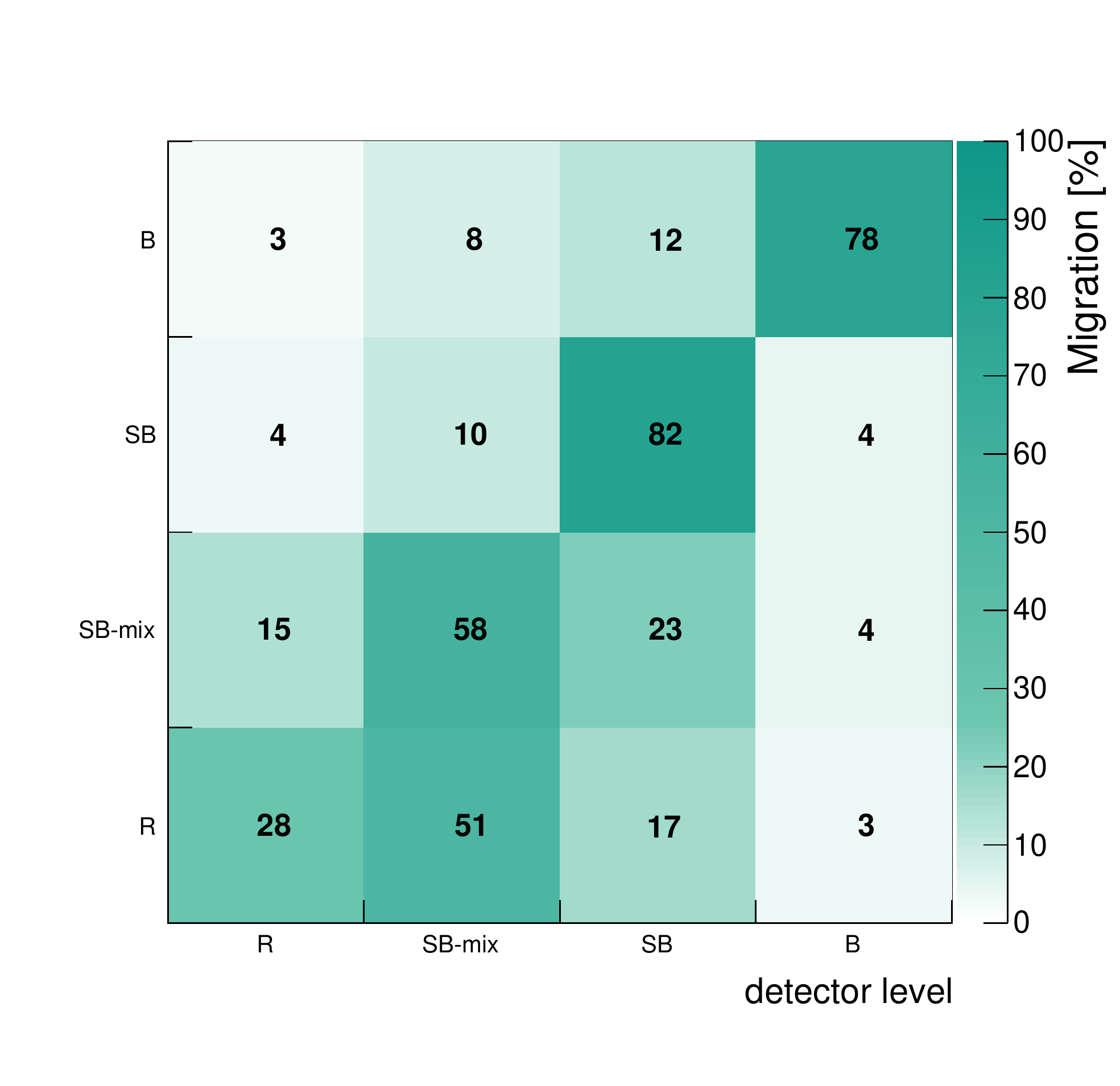}&
\includegraphics [valign=c,width=0.45\textwidth] {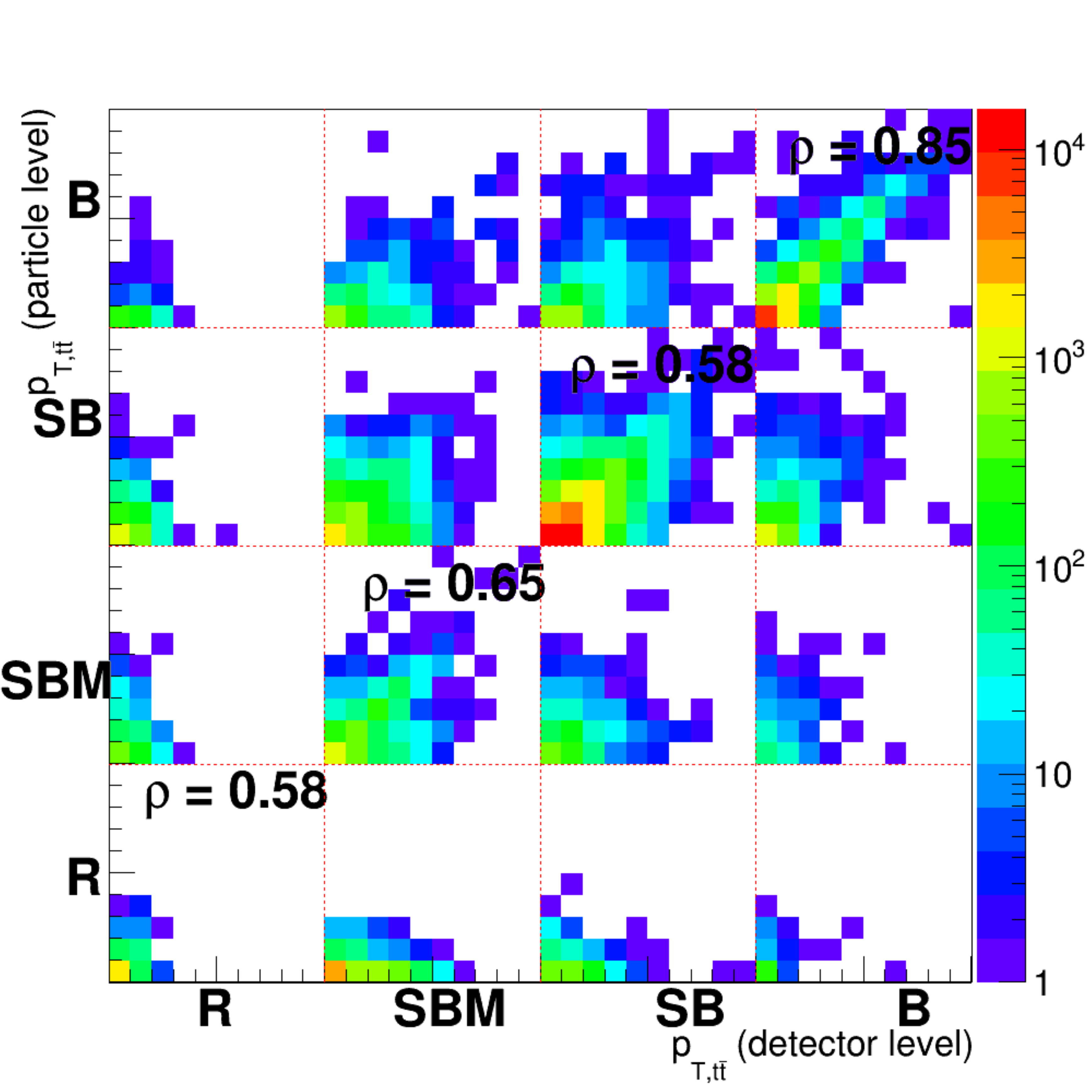} \\
\end{tabular}
\caption{The migration of events between the resolved (R), semi-boosted mixed (SB-mix, SBM), semi-boosted (SB) and boosted (B) topologies between  particle and detector levels~(left), and the  migration matrix for the transverse momentum of the reconstructed top anti-top quark pair for sample with $m_{\mathrm{Z'}}=700$~GeV~(right).}
\label{topo_mig}
\end{center}
\end{figure}

Migration of events between topologies between particle and detector levels and migration of events for a particular kinematic variable are shown in Figure \ref{topo_mig}.

The reconstruction algorithm is set to first try the boosted reconstruction, if the event does not pass the criteria, then both semi-boosted reconstruction algorithms are tried. If the event is not selected by any of them, resolved reconstruction algorithm is attempted. The event is discarded if all three reconstruction algorithms fail.


\section{Samples and resolution}

\begin{figure}
\begin{center}
\begin{tabular}{cc}
\includegraphics [width=0.45\textwidth] {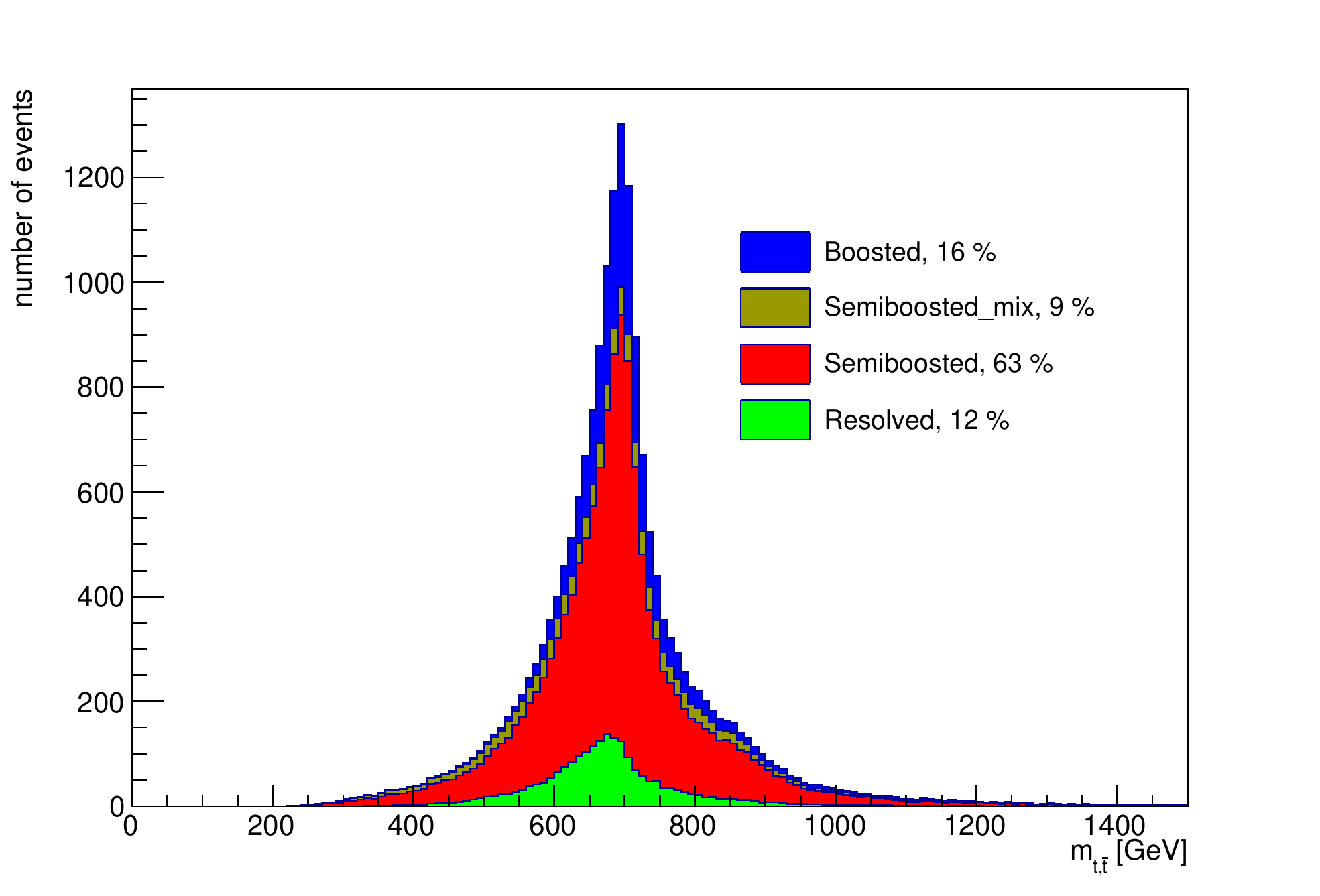}&
\includegraphics [width=0.45\textwidth] {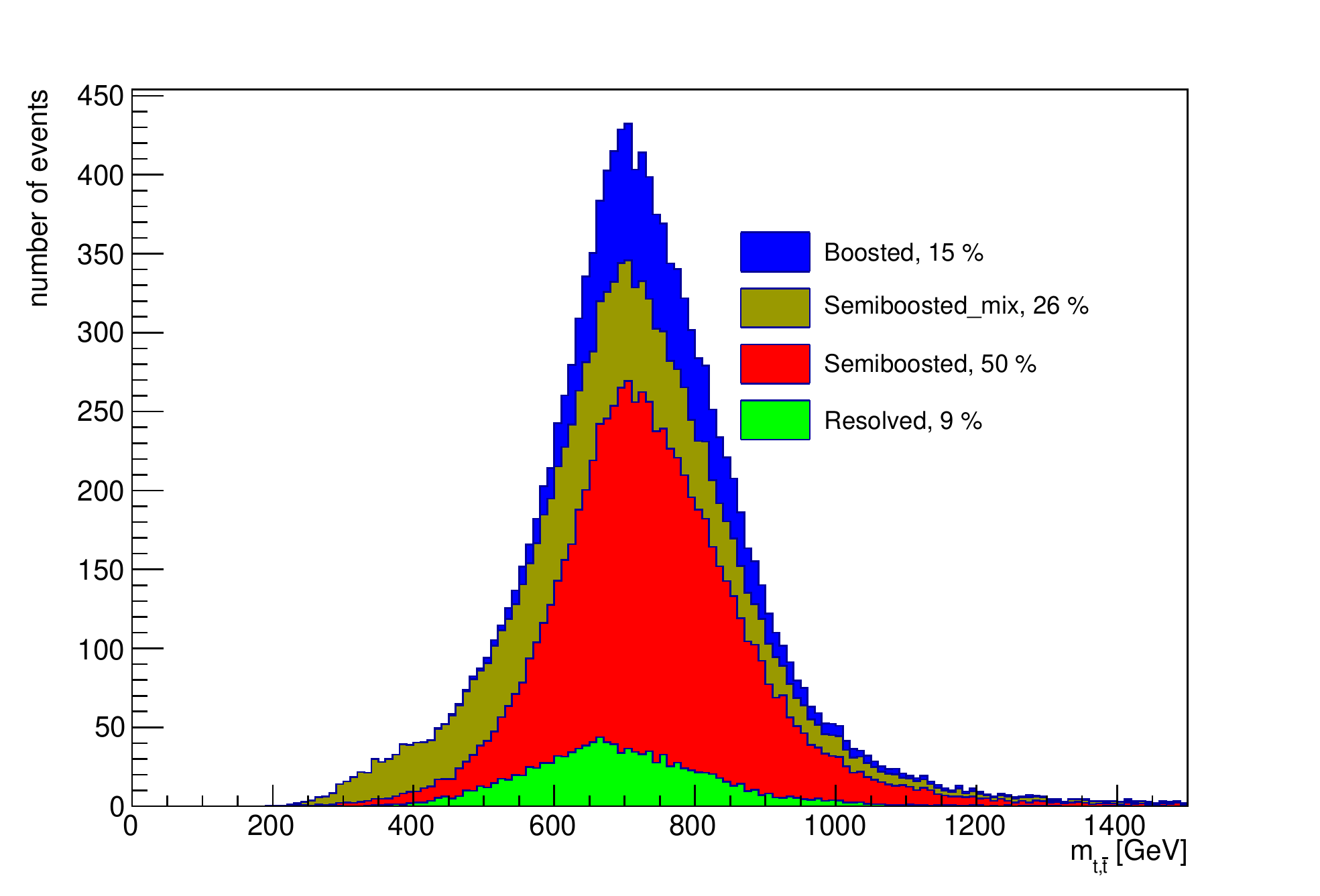}\\
\end{tabular}
\caption{Fraction of events in resolved, semi-boosted mixed, semi-boosted and boosted topologies at particle level (left) and detector level (right) for the sample with $m_{\mathrm{Z'}=700}$ GeV.}
\label{fracs}
\end{center}
\end{figure}

The simulated samples of the process $pp\rightarrow Z' \rightarrow t\bar{t}$ at $\sqrt{s} = 14$ TeV were prepared by the MadGraph generator version 2.6.4 \cite{Alwall:2014hca} interfaced with Pythia8 \cite{Sj_strand_2015} for the parton showering and hadronization simulation; and then processed by the Delphes package version 3.4.1 \cite{deFavereau:2013fsa} for a simplified parameterized detector simulation with different values of $Z'$ mass to cover the effective range of energies, where the semi-boosted topology is relevant. This range begins around 500 GeV and ends around 1000 GeV. The prepared samples were used as input for a private framework based on ROOT 6.16 \cite{Brun:1997}. Fraction of events as function of the reconstructed mass of the top anti-top quark pair for all the topologies for a sample with $m_{\mathrm{Z'}} = 700$ GeV is shown in Figure~\ref{fracs}. 

\begin{figure}
\begin{center}
\begin{tabular}{cc}
\includegraphics [width=0.45\textwidth] {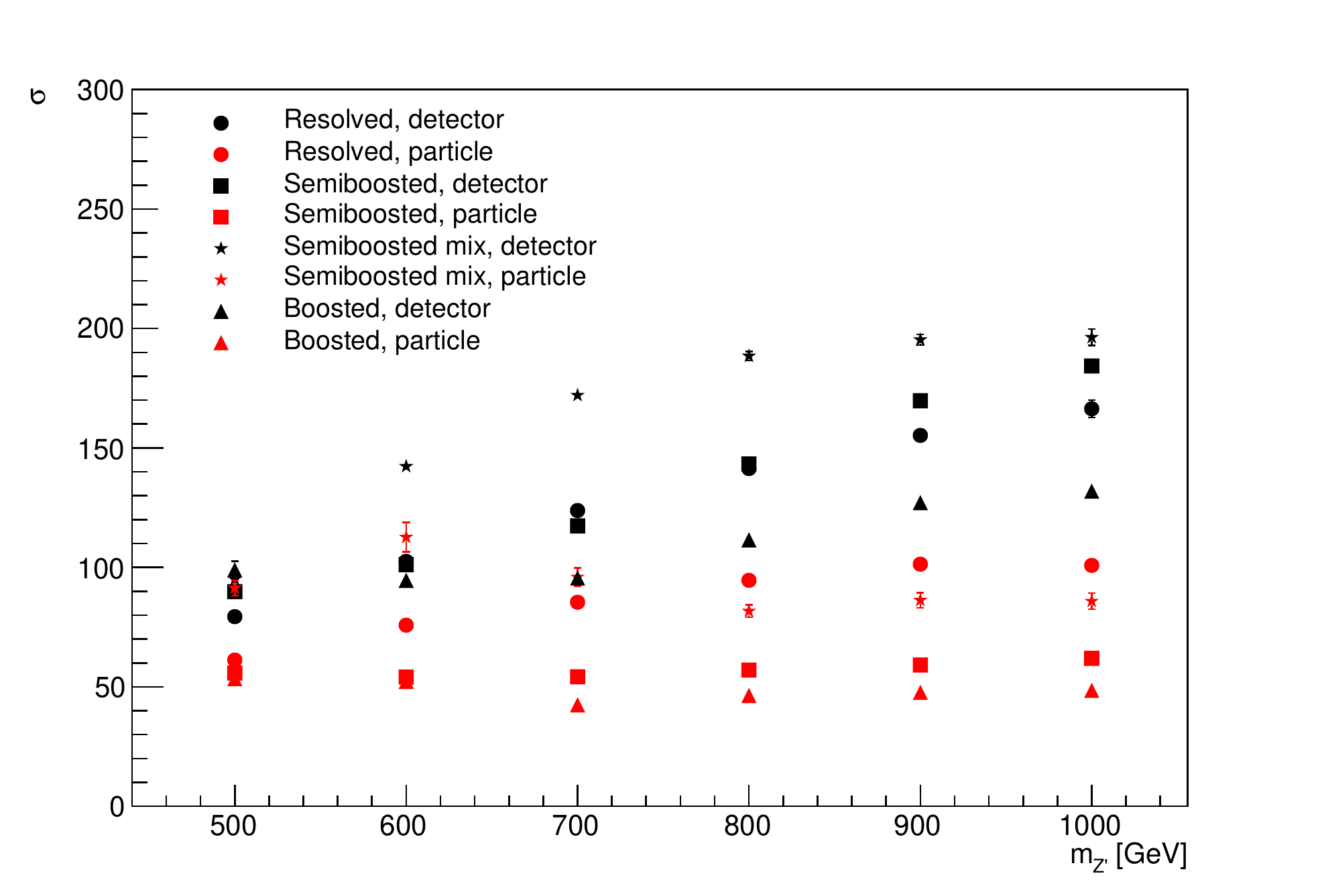}&
\includegraphics [width=0.45\textwidth] {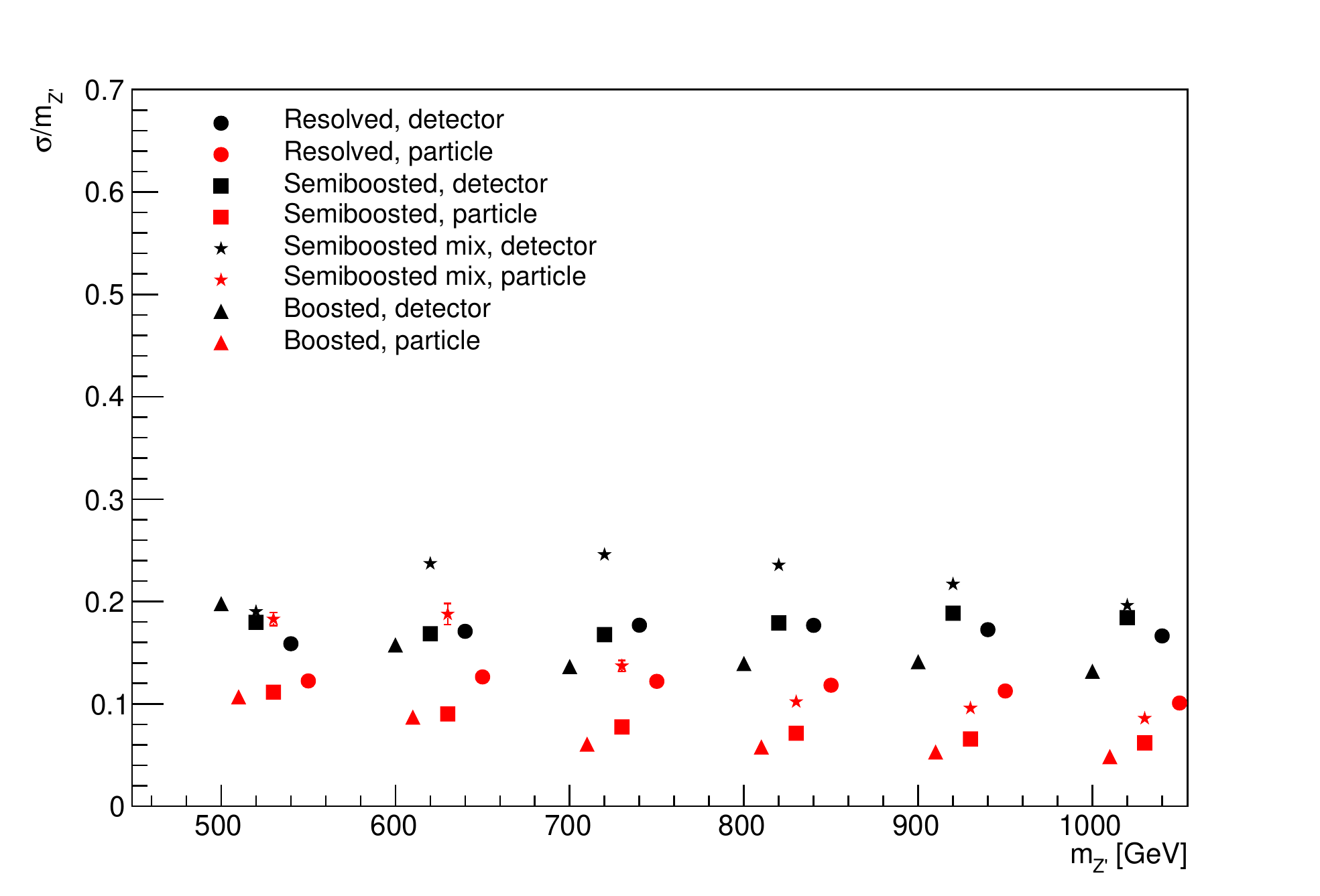}\\
\end{tabular}
\caption{Resolution of reconstructed $m_{\mathrm{t\bar{t}}}$ for topologies as function of the $Z'$ mass.}
\label{resol}
\end{center}
\end{figure}

There is an obvious difference between the resolution at the particle and detector levels. The invariant mass of the reconstructed top anti-top quark pair mass was fitted by a Gaussian function and the resolution, in terms of fitted $Z'$ mass width, for different samples and topologies is shown as both absolute and relative values with respect to $Z'$ mass of the studied sample in Figure \ref{resol}. The relative resolution in the semi-boosted case, especially at the detector level, is similar to the resolution of resolved and boosted topologies.

\section{Unfolding, closure test and the migration matrix properties}

\begin{figure}
\begin{center}
\begin{tabular}{cc}
\includegraphics[valign=c,width=0.45\textwidth]{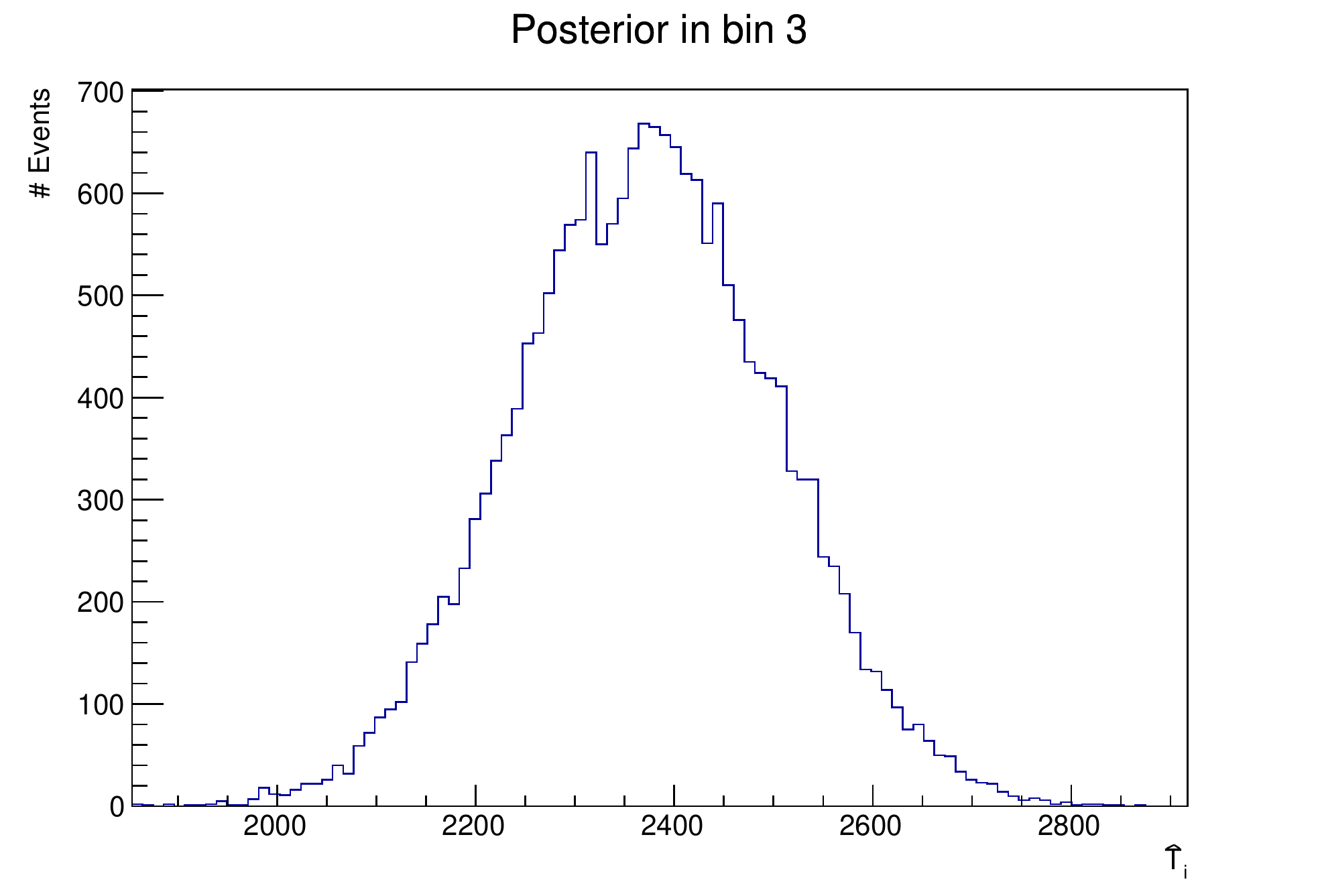}  &
\includegraphics[valign=c,width=0.45\textwidth]{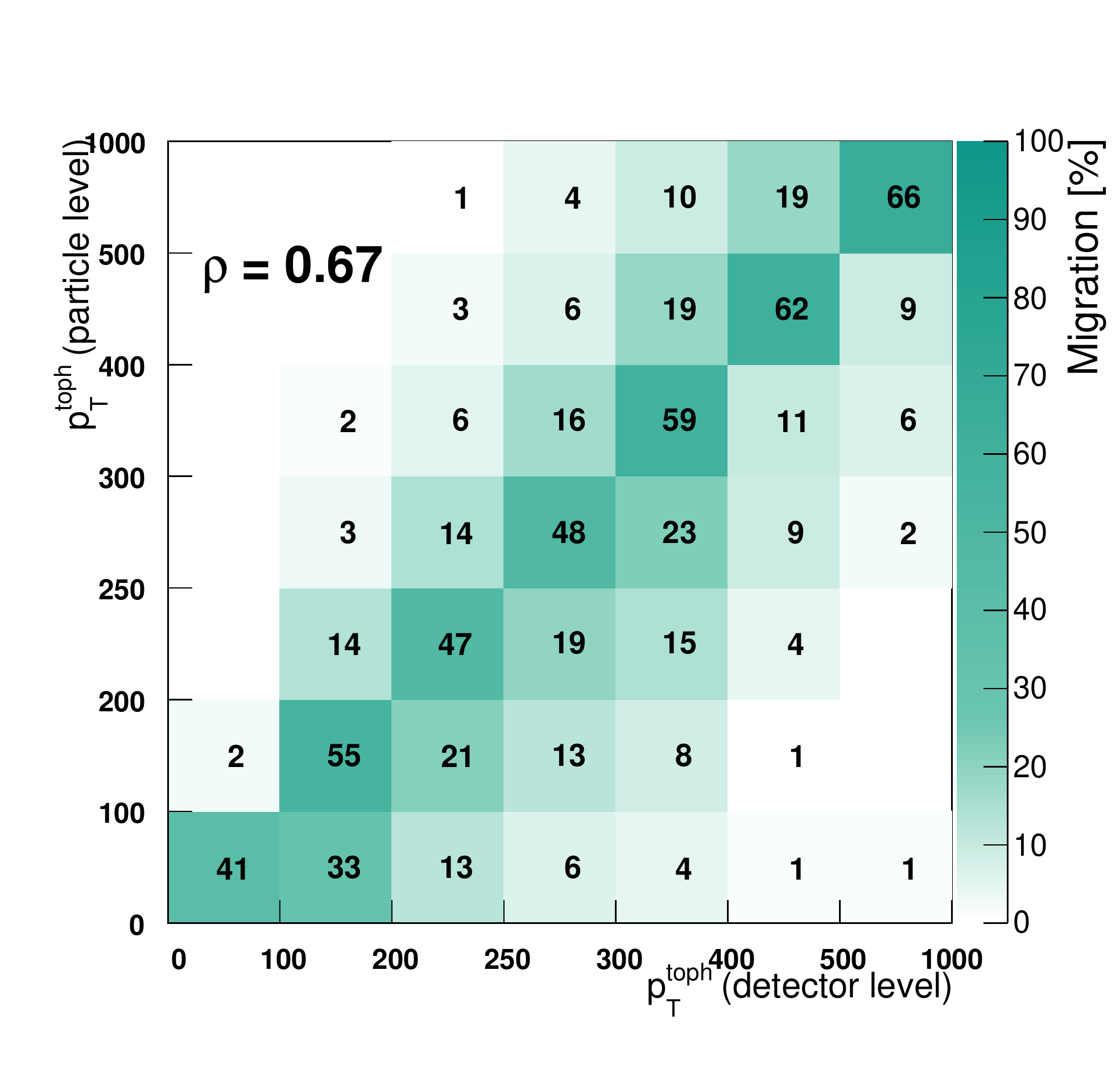}  \\
\end{tabular}
\begin{tabular}{c}
\includegraphics[width=0.65\textwidth]{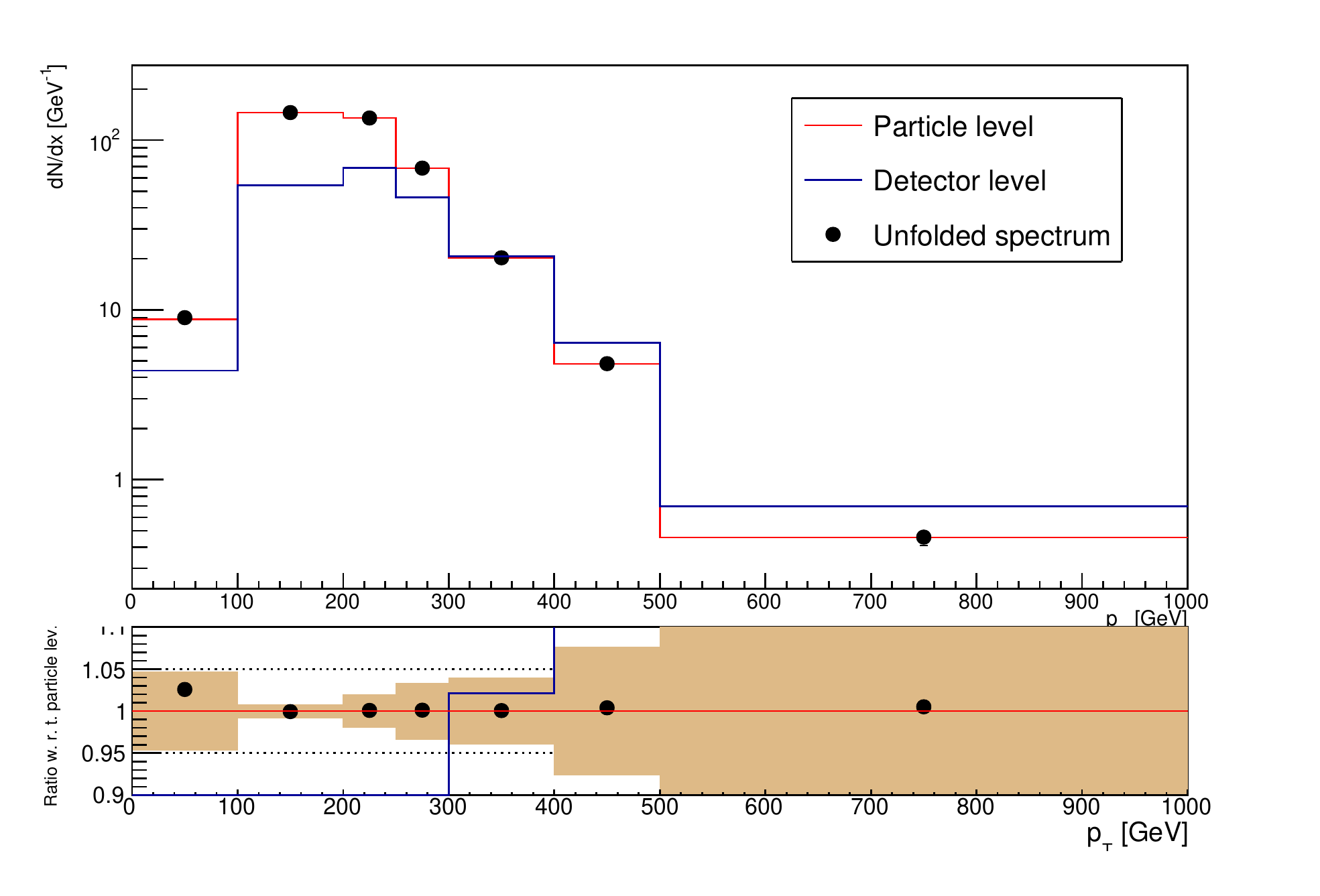} \\
\end{tabular}
\caption{An example of the posterior in a selected bin (top left), the migration matrix (top right) between the detector and the particle level; and the unfolded spectrum for $p_{\mathrm{T,toph}}$ (bottom) for the sample with $m_{\mathrm{Z'}}= 700$ GeV for the semi-boosted topology, where the ratio is with respect to the particle level spectrum and the yellow band is RMS of the posterior in the given bin.}
\label{unfolding}
\end{center}
\end{figure}

Reconstruction was also put to test through the unfolding procedure. The Fully Bayesian Unfolding (FBU) procedure \cite{choudalakis2012fully} was used to correct for influences of the detector. Unfolding can be described by the following formula  
\begin{equation}
\hat{T}_{\mathrm{i}}=f_{\mathrm{i,eff}}M_{\mathrm{ij}}^{-1}f_{\mathrm{j,acc}}(D_{\mathrm{j}} - B_{\mathrm{j}}),
\end{equation}
where $\hat{T}_{\mathrm{i}}$ is the unfolded particle level spectrum, $D_{\mathrm{j}}$ is the detector level spectrum, $B_{\mathrm{j}}$ is the background spectrum contributing to the detector level spectrum, $M_{\mathrm{ij}}$ is the corresponding migration matrix between detector and particle levels and the inversion symbolizes the unfolding procedure; and $f_{\mathrm{j,acc}}$ and $f_{\mathrm{i,eff}}$ are acceptance and efficiency correction factors. The example of the unfolded posterior, migration matrix and example of the detector, particle and unfolded spectra are shown in Figure~\ref{unfolding}. 

The unfolding procedure is working well, this can be seen from the achieved closure test in the Figure \ref{unfolding}, where the deviation of the unfolded spectrum from the particle level spectrum is under 3\%.

\section{Conclusion}
This contribution presented a study of possible efficiency enhancement of top anti-top quark pair analyses by adding semi-boosted topologies in consideration. The semi-boosted resolution is similar to other topologies in given $Z'$ mass region and it helps to increase the number of events by $60 - 70$\%, this could be beneficial especially to boosted analyses. The next step will be to find out the way to combine results from different topologies. 


\section{Acknowledgements}

This article was supported by MSMT projects GACR 19-21484S and LTT-17018 and by Palacky University project IGA\_PrF\_2019\_008.

\bibliography{ms}
\bibliographystyle{unsrt}
\end{document}